\documentclass[aps,twocolumn,pra,tightenlines,floatfix,superscriptaddress]{revtex4-1}

\usepackage[breaklinks=true]{hyperref}
\usepackage{graphicx}
\usepackage[english]{babel}
\usepackage{amsmath}
\usepackage{amssymb}
\usepackage{times}

\begin{document}

\newcommand{\D}{\mathrm{D}}
\newcommand{\p}{\partial}
\newcommand{\Tr}{\mathrm{Tr}}
\renewcommand{\d}{\mathrm{d}}
\newcommand{\Ek}{E_\mathbf{k}}
\newcommand{\xik}{\xi_\mathbf{k}}
\newcommand{\sumk}{\sum_\mathbf{k}}
\newcommand{\ek}{\xi_\mathbf{k}}
\newcommand{\ekq}{\xi_{\mathbf{k}-\mathbf{q}}}
\newcommand{\mb}[1]{{\mathbf{#1}}}
\newcommand{\uk}{u_{\mathbf{k}}}
\newcommand{\vk}{v_{\mathbf{k}}}
\newcommand{\Omegaq}{\Omega_{\mathbf{q}}}
\newcommand{\Gammaq}{\Gamma_{\mathbf{q}}}

\title{Superfluidity and pairing phenomena in ultracold atomic Fermi
  gases in one-dimensional optical lattices, Part I: Balanced case}

\author{Jibiao Wang}
\affiliation{Laboratory of Quantum Engineering and Quantum Metrology, School of Physics and Astronomy, Sun Yat-Sen University (Zhuhai Campus), Zhuhai, Guangdong 519082, China}
\affiliation{Department of Physics and Zhejiang Institute of Modern Physics, Zhejiang University, Hangzhou, Zhejiang 310027, China}

\author{Leifeng Zhang}
\affiliation{Department of Physics and Zhejiang Institute of Modern Physics, Zhejiang University, Hangzhou, Zhejiang 310027, China}

\author{Yi Yu }
\affiliation{College of Chemical Engineering, Zhejiang University of Technology, Hangzhou, Zhejiang 310014, China}

\author{Chaohong Lee}
\affiliation{Laboratory of Quantum Engineering and Quantum Metrology, School of Physics and Astronomy, Sun Yat-Sen University (Zhuhai Campus), Zhuhai, Guangdong 519082, China}
\affiliation{State Key Laboratory of Optoelectronic Materials and Technologies, Sun Yat-Sen University (Guangzhou Campus), Guangzhou, Guangdong 510275, China}

\author{Qijin Chen}
\email[Corresponding author: ]{qchen@zju.edu.cn}
 \affiliation{Shanghai Branch, National Laboratory for Physical Sciences at Microscale and Department of Modern Physics, University
  of Science and Technology of China, Shanghai 201315, China}
\affiliation{Department of Physics and Zhejiang Institute of Modern Physics, Zhejiang University, Hangzhou, Zhejiang 310027, China}
\affiliation{Synergetic Innovation Center of Quantum Information and Quantum Physics, Hefei, Anhui 230026, China}

\date{\today}

\begin{abstract}

  The superfluidity and pairing phenomena in ultracold atomic Fermi
  gases have been of great interest in recent years, with multiple
  tunable parameters. Here we study the BCS-BEC crossover behavior of
  balanced two-component Fermi gases in a one-dimensional optical
  lattice, which is distinct from the simple three-dimensional (3D)
  continuum and a fully 3D lattice often found in a condensed matter
  system. We use a pairing fluctuation theory which includes
  self-consistent feedback effects at finite temperatures, and find
  widespread pseudogap phenomena beyond the BCS regime.  As a
  consequence of the lattice periodicity, the superfluid transition
  temperature $T_c$ decreases with pairing strength in the BEC regime,
  where it approaches asymptotically $T_c = \pi an/2m$, with $a$ being
  the $s$-wave scattering length, and $n$ ($m$) the fermion density
  (mass). In addition, the quasi-two dimensionality leads to fast
  growing (absolute value of the) fermionic chemical potential $\mu$
  and pairing gap $\Delta$, which depends exponentially on the ratio
  $d/a$. Importantly, $T_c$ at unitarity increases with the lattice
  constant $d$ and hopping integral $t$. The effect of the van Hove
  singularity on $T_c$ is identified. The superfluid density exhibits
  $T^{3/2}$ power laws at low $T$, away from the extreme BCS limit.
  These predictions can be tested in future experiments.

\end{abstract}


\maketitle

\section{Introduction}

Ultracold atomic Fermi gases loaded in optical lattices have attracted
enormous attention in condensed matter and atomic, molecular and
optical (AMO) physics
\cite{Michael05PRL,Bloch_NP,GeorgesVarenna}. With multiple easily
tunable parameters, they become more and more important as a quantum
simulator nowadays \cite{Review,Bloch_RMP,Stringari_RMP}. Fermions in
\emph{pure} optical lattices are often described by a Hubbard model
\cite{Bloch_RMP,Stringari_RMP,Micnas14AP}. Among them, the
one-dimensional (1D) case can be solved exactly via the Bethe ansatz
\cite{GuanRMP}. However, while for a 1D Hubbard model, each site has
at most two fermions, the 1D optical lattice is actually rather
different; each site in the lattice direction corresponds to a 2D
plane in the transverse dimensions, and thus can accommodate many
fermions. Therefore, a 1D optical lattice is a quasi-2D or 3D system
\cite{Dyke11PRL,Martin12PRL}, depending on the lattice
parameters. Moreover, the genuine 1D Hubbard model does not possess a
long-range order, hence it can not support a superfluid phase. In
contrast, fermions trapped in 1D optical lattices can not only form a
superfluid \cite{Ries15PRL,Murthy15PRL}, but also exhibit 
interesting pseudogap phenomena in the normal state
\cite{Feld11Nature}. A condensed matter analogue of the 1D optical
lattice is the superlattice of semiconductor heterostructures such as
the AlGaAs/GaAs/InGaAs structure, except now we are considering
pairing phenomena under a tunable attractive interaction.

Including the Hubbard model, there has been extensive literature on 3D
(and 2D or 1D) lattices in the field of condensed matter
\cite{MicnasRMP,Chen1}.  Most of these existing Hubbard model based
works address pure lattice cases, since the kinetic energy term often
contains only the lattice site hopping
\cite{Demler2,*Vincent04PRA,*Michael05PRL,*Cazalilla05PRL,*Torma06NJP,*MoreoPRL98,Chien08PRA_RC,*Chien08PRA,Iskin08PRA,*WuCJPRA83,*Gottwald08EPJB,*ChenYan09PRB,*WangPRA79,*Loh10PRL,*CuiPRA81,ChenAHai2012,*Mendoza2013,Micnas14AP,Torma14PRL,*Peters15PRB,*Kitamura16PRB}. The
``1D optical lattice'' in many theoretical works in the literature was
actually a genuine 1D lattice in the traditional sense
\cite{Gu07PRB,*FeiguinPRB76,*Rizzi,*TormaPRL101,*Roscilde12EPL,*Buchleitner2012}. Here,
following the terminology often used by the experimental community
\cite{Bloch_NP}, \emph{we emphasize that, by 1D optical lattice, we mean a
periodic stack of 2D planes}, and therefore it is a mix of continuum in
the transverse 2D $xy$ planes and lattice discreteness in the
longitudinal $z$ direction.  Theoretical studies on such a 1D optical
lattice have been scarce. Devreese \textit{et al.} studied possible
Fulde-Ferrell-Larkin-Ovchinnikov (FFLO) states \cite{FF,LO} in such a
1D optical lattice
\cite{DevreesePRA83,*Devreese2011,*Devreese12MPLB}. Like many others
in the literature \cite{Micnas14AP}, when studying population
imbalance effects, they use the fermion chemical potential $\mu$ and
the chemical potential difference $h$ as control variables. While this
choice makes numerical calculations simpler, it often restricts the
study to the BCS and crossover regimes.
Indeed, the superfluid and pairing physics in a 1D optical lattice has
not been adequately studied thus far. Given the various available
tuning parameters, including the pairing interaction strength, lattice
constant and depth, fermion density, population imbalance, as well as
mass imbalance in the case of a Fermi-Fermi mixture, there are
certainly many facets of the phase diagram and associated very rich
physics. In particular, one would like to know if there are exotic new
phases emerging, and how to properly characterize such a 1D optical
lattice.

In this paper, we study two-component fermions loaded in 1D optical
lattices using a pairing fluctuation theory, which has been applied
successfully to various BCS-BEC crossover phenomena
\cite{Review,ChenPRL98,chen07prb,OurRFReview}, including in quasi-2D
and 3D optical lattices
\cite{Chen1,Chien08PRA_RC,*Chien08PRA,Chen12PRA}. These systems can be
quasi-2D or 3D, depending on the lattice constant $d$ and hopping
integral $t$ \cite{Zhang17SR}, as well as the pairing strength. Due to
the complexity induced by multiple tunable parameters, in this paper
(Part I), we restrict ourselves to population (and mass) balanced
cases only. Here we consider the combined effects of lattice constant,
hopping integral, and interaction strength. We find that the mixing
between continuum and discrete lattice dimensions leads to exponential
behavior of the fermionic chemical potential $\mu$ and the pairing gap
$\Delta$ as a function of $d/a$ in the BEC regime, where $a$ is the
two-body $s$-wave scattering length, in contrast to the power laws in
the pure 3D continuum or 3D lattice cases.  We shall present detailed
phase diagrams as the system undergoes the BCS-BEC crossover with
different lattice constants and hopping integrals, and mainly focus on
the finite temperature effects, especially the pseudogap phenomena
\cite{Chen14Review,Mueller17RPP}.  As these phase diagrams reveal, (i)
the pseudogap phenomena widely exist; (ii) At unitarity, $T_c$
increases with the increase of lattice constant or hopping integral;
(iii) As a consequence of the lattice periodicity, $T_c$ decreases
with pairing strength in the BEC regime, and approaches asymptotically
$T_c = \pi an/2m$, where $n$ is the atom number density, $m$ the atomic
mass;
(iv) In addition, the quasi-two dimensionality leads to fast growing
(absolute value of the) fermionic chemical potential $\mu$ and pairing
gap $\Delta$, which depends exponentially on the ratio $d/a$; (v) Due
to the contribution of finite momentum pairs, the temperature
dependence of the superfluid density $n_s/m$ at low $T$ evolves from
exponential in the extreme BCS limit to a simple $T^{3/2}$ power law in
the BEC regime, for both the in-plane and the out-of-plane (lattice)
components.  
  
\section{Theoretical Formalism}

\subsection{General theory}

While the in-plane ($xy$ directions) motion of the fermions has a free
parabolic dispersion, we use a one-band nearest-neighbor tight-binding
model for the out-of-plane lattice dimension ($z$ direction), with the
single particle dispersion given by
$\xi_{\textbf{k}\sigma}=\textbf{k}^{2}_{\parallel}/2m+2t[1-\cos(k_{z}d)]-\mu_{\sigma}\equiv
\epsilon_{\textbf{k}}-\mu_{\sigma}$. Here
$\textbf{k}_{\parallel}\equiv(k_x,k_y)$ and $t$ is the hopping
integral between neighboring lattice sites, $d$ the optical lattice
constant and $\mu_{\sigma}$ the fermionic chemical potentials for two
(pseudo)spins $\sigma=\uparrow,\downarrow$. In the absence of
imbalance, we have $\mu_\sigma = \mu$, and
$\xi_{\textbf{k}\sigma}^{} = \xik^{}$, and we shall drop the spin
indices. We restrict $k_{z}$ to the first Brillouin zone (BZ)
$[-\pi/d,\pi/d]$ due to the lattice periodicity and set the volume
$V=1$, $\hbar=k_{B}=1$. The one-band approximation is justified when
the band gap is tuned large. The fundamental formalism of the pairing
fluctuation theory for the present work is the same as that given in
Refs. \cite{ChenPRL98,Review}, except that we need to rederive the
equations with the continuum-lattice mixed dispersion. To keep this
paper self-contained, here we recapitulate the derivation and main
equations.

The (inverse) bare Green's function is given by
$G^{-1}_{0}(K)=i\omega_{n}-\xi_{\textbf{k}}$, with the self-energy
$\Sigma(K)=\sum_{Q}t(Q)G_{0}(Q-K)$. Following Ref.~\cite{ChenPRL98},
we use a four-vector notation,
$\sum_{K}\equiv T\sum_{n}\sum_{\textbf{k}}$,
$\sum_{Q}\equiv T\sum_{l}\sum_{\textbf{q}}$, and
$K\equiv(i\omega_{n},\mathbf{k})$, $Q\equiv (i\Omega_{l},\textbf{q})$,
where $\omega_{n}=(2n+1)\pi T$, $\Omega_{l}=2l\pi T$ are odd and even
Matsubara frequencies, respectively \cite{Fetter}. At finite $T$, the
$T$-matrix $t(Q)$ contains a contribution from condensed pairs
$t_{sc}(Q)$ and noncondensed pairs $t_{pg}(Q)$, with
$t(Q)=t_{sc}(Q)+t_{pg}(Q)$, where
$t_{sc}(Q)=-(\Delta_{sc}^{2}/T)\delta(Q)$ vanishes for $T>T_c$, and
$t_{pg}(Q)=U/[1+U\chi(Q)]$, with the short range $s$-wave pairing
interaction pairing strength $U<0$ and the pair susceptibility
$\chi(Q)=\sum_{K}G_{0}(Q-K)G(K)$.  Here $G(K)$ is the full Green's
function, with the self-energy given by
$\Sigma(K)=\Sigma_{sc}(K)+\Sigma_{pg}(K)$, where
$\Sigma_{sc}(K)=\sum_{Q}t_{sc}(Q)G_{0}(Q-K)=
-\Delta_{sc}^{2}G_{0}(-K)$, and
$\Sigma_{pg}(K)=\sum_{Q}t_{pg}(Q)G_{0}(Q-K)$. At $T\le T_c$, the
generalized Thouless criterion \cite{Thouless}, or equivalently BEC
condition for pairs, requires
$t^{-1}_{pg}(Q=0)=U^{-1}+\chi(0)=0$. This implies that $t_{pg}(Q)$ is
dominated by the vicinity of $Q=0$, so that $\Sigma_{pg}(K)$ may be
approximated by
$\Sigma_{pg}(K)\approx\sum_{Q}t_{pg}(Q)G_{0}(-K)\equiv-\Delta_{pg}^{2}G_{0}(-K)$,
where $\Delta_{pg}^{2}\equiv-\sum_{Q}t_{pg}(Q)$ and we have discarded
the incoherent background part of the self energy. (The parameter
$\Delta_{pg}$ is referred to as pseudogap, as is widely found in
cuprate superconductors \cite{Timusk}). Then the total self-energy
$\Sigma(K)$ takes the simple BCS-like form,
$\Sigma(K)=-\Delta^{2}G_{0}(-K)$, where
$\Delta^{2}=\Delta_{sc}^{2}+\Delta_{pg}^{2}$. Finally, the Dyson's
equation $G^{-1}(K)=G^{-1}_{0}(K)-\Sigma(K)$ leads immediately to the
full Green's function
\begin{equation}
 G(K)=\frac{u_{\textbf{k}}^{2}}{i\omega_{n}-E_{\textbf{k}}}+\frac{v_{\textbf{k}}^{2}}{i\omega_{n}+E_{\textbf{k}}}\,,
\end{equation}
where $u_{\textbf{k}}^{2}=(1+\xi_{\textbf{k}}/E_{\textbf{k}})/2$,
$v_{\textbf{k}}^{2}=(1-\xi_{\textbf{k}}/E_{\textbf{k}})/2$, and
$E_{\textbf{k}}=\sqrt{\xi_{\textbf{k}}^{2}+\Delta^{2}}$. From the
number constraint $n=2\sum_{K}G(K)$, we can get the fermion number
density
\begin{eqnarray}
 n&=&2\sum_{\textbf{k}}\Big[v_{\textbf{k}}^{2} + {f}(E_{\textbf{k}})\frac{\xi_{\textbf{k}}}{E_{\textbf{k}}}\Big]\,,\label{eq:LOFF_neqa}
\end{eqnarray}
where $f(x) = 1/(e^{x/T}+1)$ is the Fermi distribution function.

Above $T_c$, the Thouless criterion should be modified by
$U^{-1}+\chi(0)=a_{0}\mu_{p}$, where $\mu_{p}$ is the effective pair
chemical potential and $a_{0}$ is the coefficient of the linear
$\Omega$ term in the Taylor expansion of the inverse $T$-matrix (see
below) \cite{Review}. This leads to the extended gap equation
\begin{equation}
  \frac{m}{4\pi a}=\sum_{\textbf{k}}\Big[\frac{1}{2\epsilon_{\textbf{k}}}-\frac{1-2{f}(E_{\textbf{k}})}{2E_{\textbf{k}}}\Big]+a_{0}\mu_{p}\,,
  \label{eq:gap}
\end{equation}
with $\mu_{p}=0$ at $T\leq T_{c}$. Here, the coupling strength $U$ has
been replaced by the $s$-wave scattering length $a$ via
$U^{-1}=m/4\pi a-\sum_{\textbf{k}}1/2\epsilon_{\textbf{k}}$. Note that
this scattering length is different from that defined in simple 3D
free space, since $k_z$ is now restricted to within the first BZ. We
caution that it does \emph{not} necessarily yield the experimentally
measured scattering length. One can define an effective scattering
length via $a_\text{eff} = a/\sqrt{2mtd}$, which is more comparable to
the physical scattering length. For details, see Ref.~\cite{1DOLshort}
and its supplementary materials.  It should also be noted that we have
implicitly assumed a negative $U$ model for the lattice direction and
the on-site $U$ is same as the in-plane pairing strength $U$. In real
space, the pairing interaction is given by
$U(\mathbf{r},\mathbf{r}') = U \delta(x-x')\delta(y-y')\delta_{ij}$,
where $i,j$ is the lattice site index in the $\hat {z}$ direction.

The inverse $T$-matrix expansion \cite{Review}, after analytic
continuation $(i\Omega_{l}\rightarrow\Omega+i0^{+})$, is given by
\begin{equation}
t_{pg}^{-1}(\Omega,\textbf{q})\approx
a_{1}\Omega^{2}+a_{0}(\Omega-\Omega_{\textbf{q}}+\mu_{p}),
\label{eq:Texpansion}
\end{equation}
with
$\Omega_{\textbf{q}}=B_{\parallel}\textbf{q}_{\parallel}^{2}+2t_{B}[1-\cos(q_{z}d)]$. Here
$B_{\parallel}=1/2M_\parallel$, with $M_\parallel$ being the effective
pair mass in the $xy$-plane, and $t_{B}$ is the effective hopping
integral for noncondensed pairs. The $a_1$ term serves as a small
quantitative correction; except in the weak coupling BCS regime, we
have $a_1T_c \ll a_0$. The coefficients $a_{1}$, $a_{0}$,
$B_{\parallel}$ and $t_{B}$ can be derived from the pair
susceptibility via straightforward Taylor expansion, as given in the
Appendix. Consequently, we have the pseudogap equation
\begin{equation}
 a_0\Delta_{pg}^{2}=\sum_{\textbf{q}}\frac{b(\tilde{\Omega}_{\textbf{q}})}{\sqrt{1+4\dfrac{a_{1}}{a_{0}}(\Omega_{\textbf{q}}-\mu_{p})}}\,,\label{eq:PG}
\end{equation}
where $b(x)$ is the Bose distribution function and
$\tilde{\Omega}_{\textbf{q}}=\{\sqrt{a_{0}^{2}[1+4a_{1}(\Omega_{\textbf{q}}-\mu_{p})/a_{0}]}-a_{0}\}/2a_{1}$
is the pair dispersion. When $a_1/a_0$ is small, we have
$\tilde{\Omega}_\mathbf{q} =\Omega_{\textbf{q}}-\mu_{p}$.  Then
$a_0\Delta_{pg}^2$ yields the density of finite momentum
pairs. Including the condensate, the total pair density is given by
$n_p = a_0\Delta^2$.

Equations (\ref{eq:LOFF_neqa})-(\ref{eq:PG}) form a closed set of
self-consistent equations, which can be used to solve for ($\mu$,
$T^*$) with $\Delta=0$, for ($\mu$, $\Delta_{pg}$, $T_c$) with
$\Delta_{sc}=0$, and for ($\mu$, $\Delta$, $\Delta_{pg}$) at
$T<T_c$. Here the pair formation temperature $T^*$ is approximated by
the mean-field $T_c$, and the order parameter $\Delta_{sc}$ can be
derived from $\Delta_{sc}^2=\Delta^2-\Delta_{pg}^2$ below $T_c$.

\subsection{Asymptotic behavior in the deep BEC regime}

In the deep BEC regime, $\mu\rightarrow -\infty$. The integrals in the
equations can be performed analytically using Taylor expansions. The
fermion number equations reduce to
\begin{equation}
  n = -\frac{m\Delta^2}{4\pi \mu d}\quad\mbox{or}\quad
  \Delta = \sqrt{\frac{4\pi|\mu|dn}{m}}\,.
  \label{eq:nBEC}
\end{equation}
With the help of Eq.~(\ref{eq:nBEC}), the chemical potential $\mu$ can
be uniquely determined by the gap equations. Then $\mu$ and the gap
$\Delta$ are given by
\begin{eqnarray}
  \label{eq:mu}
  \mu &=&  -te^{d/a} + 2t + \frac{2\pi dn}{m},\\
  \Delta &=& 2\sqrt{\frac{\pi tdn}{m} } e^{d/2a}\left(1 - \frac{\pi dn}{mt}  e^{-d/a}\right)\,.
  \label{eq:Delta}
\end{eqnarray}
Note that the exponential behavior of $\mu$ and $\Delta$ as a function
of $1/k_Fa$ is an important feature of the quasi-two dimensionality of
the continuum-lattice mixed system. This should be contrasted with the
corresponding behaviors in the 3D continuum and 3D lattices, where
power law dependencies are found. In particular, a 3D continuum has
the scaling relation $\Delta \sim |\mu|^{1/4}$ in the BEC regime and
thus $\Delta^2/\mu$ decreases with $1/k_Fa$.  On the other hand, for a
3D lattice, due to the finite volume of the unit cell, both $|\mu|$
and $\Delta$ grow linearly with $|U|$, with a ratio of
$\Delta/|\mu| = \sqrt{2n-n^2}/(1-n)$ for $n<1$ per unit cell. In
contrast, for the present continuum-lattice mixed system, the ratio
$\Delta^2/\mu $ approaches a constant, independent of pairing
strength.
For this reason, the (2nd and 3rd) correction terms in
Eq.~(\ref{eq:mu}) are also constants, independent of the interaction
strength. The correction term in Eq.~(\ref{eq:Delta}) quickly drops as
$|U|$ increases.

To solve for $T_c$, we first derive the pair dispersion, and find
\begin{eqnarray}
  \label{eq:BBEC}
  B_\parallel &=& \frac{1}{4m}, \\
  t_B &=& \frac{t^2}{2|\mu|} \approx  \frac{t}{2} e^{-d/a}\,.
\label{eq:BzBEC}          
\end{eqnarray}
While the in-plane pair mass in the BEC regime is given by $2m$, as
expected, the out-of-plane pair mass becomes exponentially heavy, as a
function of increasing $d/a$. This can be easily understood since on a
lattice pairs hop mainly via ``virtual ionization'' \cite{NSR} (i.e.,
virtual pair unbinding) and thus its mobility is inversely
proportional to the pair binding energy $2|\mu|$.  The pseudogap
equation now becomes the equation for pair density $n_p$,
\begin{equation}
  a_0\Delta^2 \equiv n_p = \frac{n}{2},
  \label{eq:a0BEC}
\end{equation}
and the coefficient $a_1$ is given by
\begin{equation}
  a_1\Delta^2 = -\frac{n}{8\mu}\,,
  \label{eq:a1BEC}
\end{equation}
which becomes exponentially small in the BEC limit.  Now
one readily derive the solution for $T_c$,
\begin{equation}
  T_c = \frac{2\pi B_\parallel dn}{d/a - \ln (t/T_c)} \approx \frac{\pi  an}{2m} = \frac{k_Fa}{3\pi}T_F\,,
  \label{eq:TcBEC}
\end{equation}
where use has been made of the definition of
$k_F^{} = (3\pi^2 n)^{1/3}$ and $E_F^{} = T_F^{} = k_F^2/2m$ (as in 3D
continuum) in the last step, and we have dropped the small logarithmic
correction $\ln (t/T_c)$
in the
denominator. An important and interesting aspect of this result is
that \emph{the BEC asymptote is essentially independent of $d$}, and
the effect of $t$ only enters through a logarithmic correction, which
can be safely neglected in the asymptote as well.

\subsection{Superfluid density}

Given the solution of the self-consistent equations, one can easily
investigate the transport behavior of the system. As an example, in
this subsection, we shall present calculations for superfluid
``density'' $n_s/m$, which is important quantity in the superfluid
phase. In superconductors, it is often measured via the London
penetration depth $\lambda_L$, especially at low $T$, with the
relation $n_s/m \propto \lambda_L^{-2}$. The temperature dependence at
low $T$ often serves as a strong indicator for the pairing symmetry of
a superconductor, as it depends strongly on the pairing symmetry.  BCS
mean-field calculations show that it exhibits exponential $T$
dependence for an $s$-wave superconductor, and linear $T$ dependence
for a nodal $d$-wave superconductor \cite{Schrieffer,Timusk,Leggett1996Review,ChenPRL98}.

The expression for superfluid density can be derived following
Refs.~\cite{Kosztin1,ChenPRL98}, using the linear response
theory. More technical details can be found in Ref.~\cite{ChenPhD}.
For the present charge-neutral atomic gases, we only need to assume a
fictitious vector potential, which can actually be realized
experimentally via synthetic gauge fields.

Without imbalance, the superfluid density is given by
\begin{equation}
  \left(\frac{n_s}{m} \right)_{i}     = 2  \sum_\mathbf{k} \frac{\Delta_{sc}^2}{\Ek^2} \left[ \frac{1-2{f}(\Ek)}{2\Ek} +{f}^\prime(\Ek)\right]
                                        \left( \dfrac{\partial\xik}{\partial{k}_i}\right)^2 ,
  \label{eq:Ns}
\end{equation}
where $i=x,y,z$ and
$f'(x) = -f(x)f(-x)/T$ is the directive of the Fermi distribution function.

Following Ref.~\cite{ChenPhD}, it can be shown that for the in-plane
motion, $(n_s/m)_\parallel = (n_s/m)_x = (n_s/m)_y  = n/m$ at $T=0$, since
$\partial^2 \xik /\partial k_i^2 = 1/m =\text{const}$ for $i=x,y$. In
contrast, in the lattice direction, the inverse band mass
$(1/m)_{z} = {\partial^2\xik}/{\partial{k}_z^2} = 2td^2 \cos(k_zd)$
is $k_z$ dependent and scaled by the factor $td^2$. As a consequence,
we expect $(n_s/m)_z \propto t^2d^2 $ and becomes small for realistic
lattices, based on Eq.~(\ref{eq:Ns}).

\section{Numerical Results and Discussions}

\subsection{Effect of lattice-continuum mixing on BCS--BEC crossover}

In this subsection, we first investigate the effect of
lattice-continuum mixing on the behavior of $T_c$ and phase diagram
throughout the BCS--BEC crossover regimes.

\begin{figure}
\centerline{\includegraphics[clip,width=3.3in]{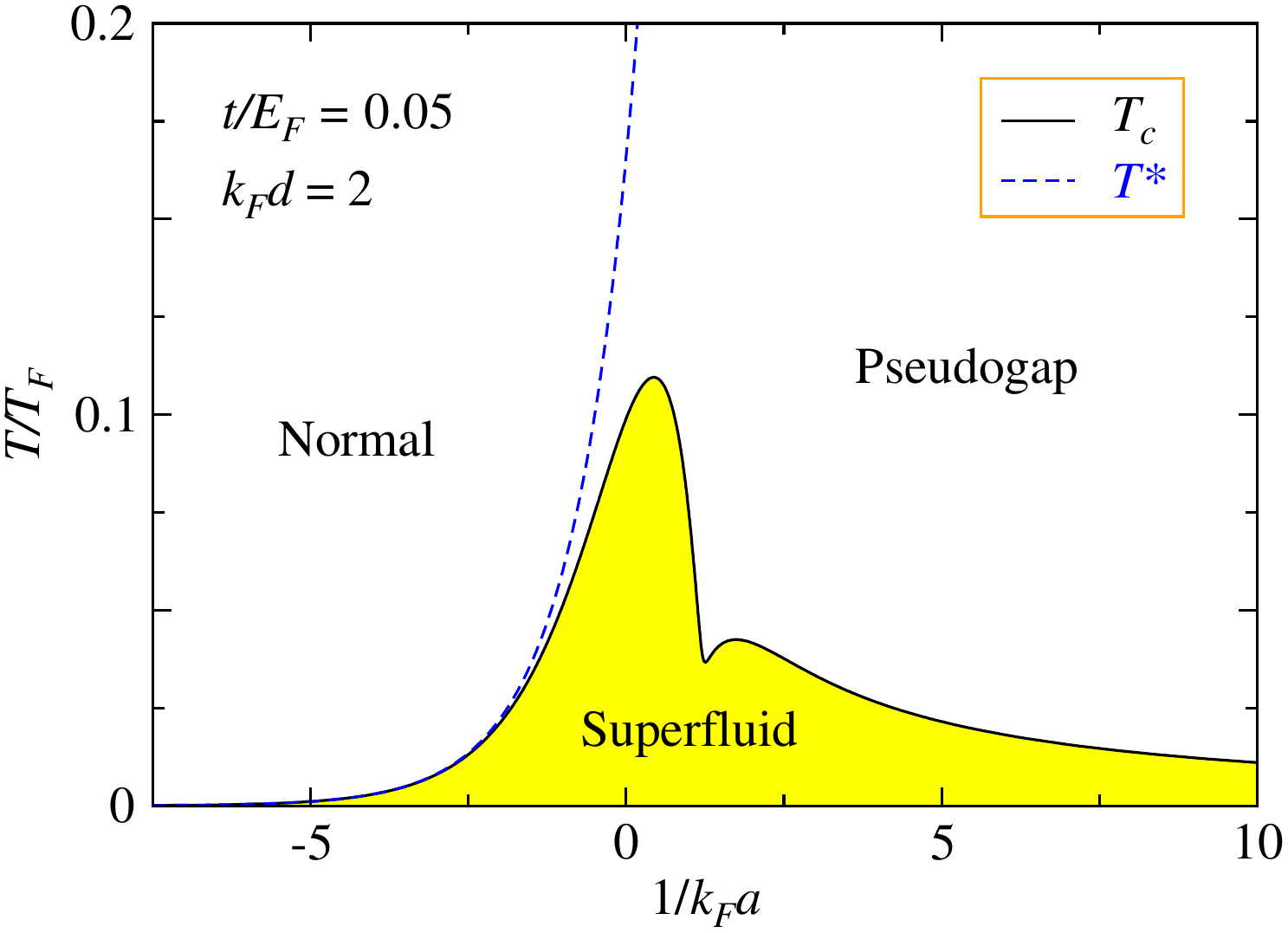}}
\caption{Typical phase diagram in the $T$ -- $1/k_Fa$ plane, calculated for
  $t/E_F=0.05$ and $k_Fd=2$.}
\label{fig:phase}
\end{figure}

Shown in Fig.~\ref{fig:phase} is a typical phase diagram of a
two-component balanced Fermi gas in a 1D optical lattice. Here we take
realistic values for $t$ and $d$, with $t/E_F = 0.05$ and $k_Fd =
2$. Note that in the zero lattice depth limit, the fermion energy in
the lattice dimension should reduce to the simple parabolic
dispersion, with mass $m$. Therefore, we set $td^2 < 1/2m$ as a
constraint on the choice of the values of $t$ and $d$. Here we have
$2mtd^2 = 0.2$. The (yellow) shaded area is the superfluid phase,
whereas the (blue) dashed curve is the mean-field solution of
$T_c$. We take this as an estimate of the pair formation temperature,
$T^*$. Between the $T^*$ and $T_c$ curves, there exists the pseudogap
phase, where incoherent pairs exist but without phase coherence or
Bose condensation. The $T_c$ curve reaches a maximum in the vicinity
of unitarity, where $1/k_Fa=0$. In the BEC regime, $T_c$ decreases
with increasing pairing strength. Note that the existence of the
pseudogap phase is an inevitable feature of the BCS-BEC crossover.

This phase diagram looks qualitatively similar to that in a 3D or
quasi-2D pure lattice \cite{Chen1,Chen12PRA}.  However, we note that
it in fact exhibits features of both pure 3D continuum and pure
lattice cases. On the one hand, there is a minimum in $T_c$ around
where the fermionic chemical potential $\mu$ changes sign, a feature
of 3D continuum \cite{Chen1}. On the other hand, the decrease of $T_c$
with increasing $1/k_Fa$ in the BEC regime is a feature of pair
hopping via virtual ionization \cite{NSR,Chen1} in a lattice. The BEC
regime is not accessible at high densities in a 3D or quasi-2D pure
lattice. In a typical 3D lattice, the minimum disappears, leaving only
a kink as a residue of the minimum \cite{Chen12PRA}. In a quasi-2D
lattice, such a minimum may exist only in the low density regime,
where the inter-particle distance becomes much larger than the lattice
constant. Indeed, the present system with an in-plane continuum space
should be comparable to the low density limit when compared to the
quasi-2D lattice case.

\begin{figure}
\centerline{\includegraphics[clip,width=3.3in]{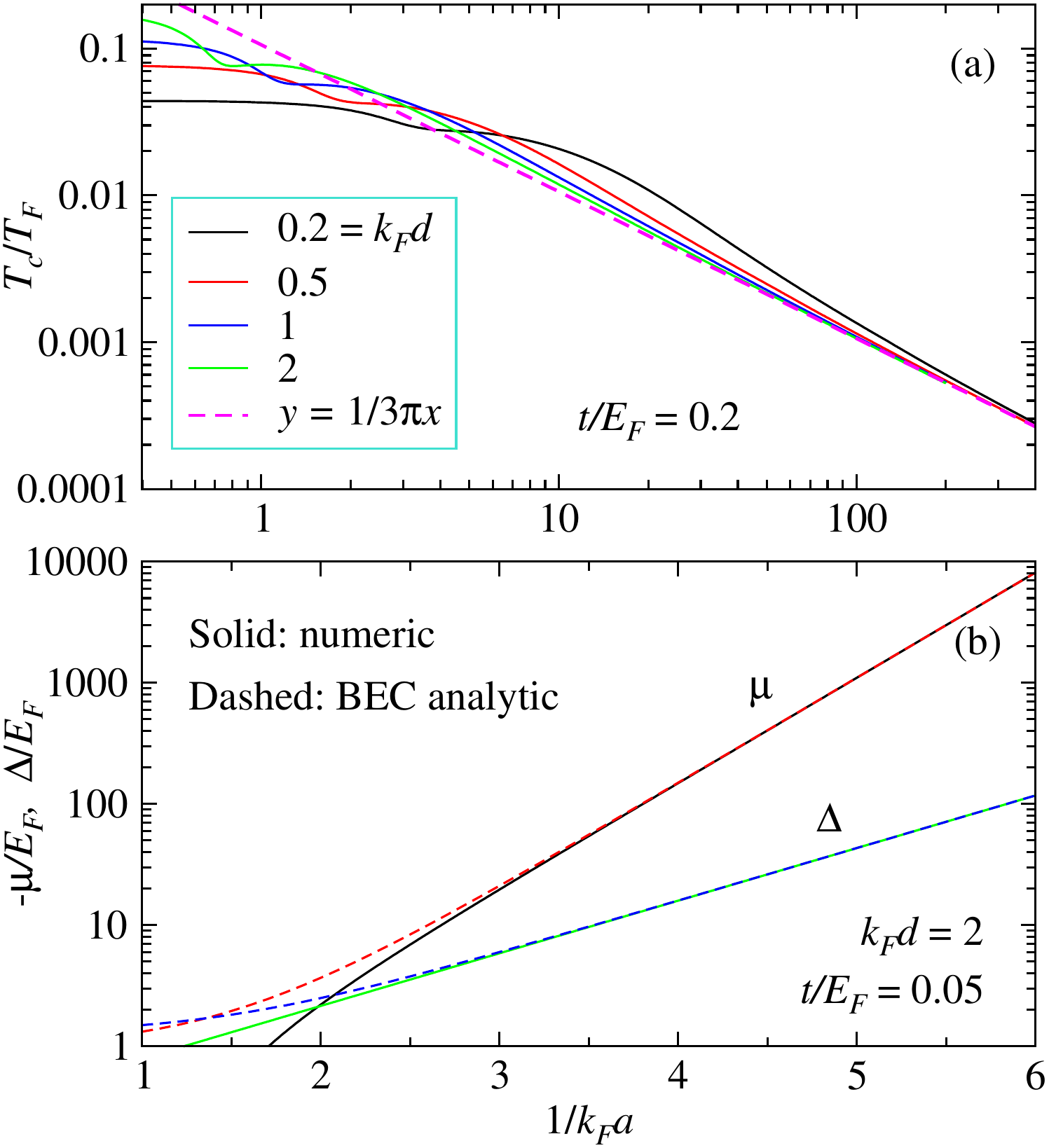}}
\caption{Comparison between fully numerical and analytical solutions
  in the BEC regime for (a) $T_c$ and (b) $\mu$, as a function of
  $1/k_Fa$. Shown in (a) are log-log plots of $T_c$ (solid lines) for
  $t/E_F=0.2$ and varying $k_Fd$ from 0.2 to 2, while the dashed line
  represents the analytical solution, Eq.~(\ref{eq:TcBEC}). Plotted in
  (b) are $-\mu$ and $\Delta$ in a semi-log scale for $t/E_F=0.05$ and
  $k_Fd=2$, where the analytical solutions (dashed lines) are given by
  Eqs.~(\ref{eq:mu}) and (\ref{eq:Delta}).}
\label{fig:BEClimit}
\end{figure}

There are further distinctions between the present lattice-continuum
mix and the pure systems. In Fig.~\ref{fig:BEClimit}, we show the
comparison between the fully numerical and analytical solutions of (a)
$T_c$ and (b) $\mu$ as a function of $1/k_Fa$ in the BEC regime. Shown
in Fig.~\ref{fig:BEClimit}(a) are the $T_c$ curves in log-log scales
for different values of $d$, while keeping $t$ fixed at
$t=0.2E_F$. Also shown is the analytically solution,
Eq.~(\ref{eq:TcBEC}), in the BEC regime (magenta dashed line). As is
evident, all $T_c$ curves approach this $(t,d)$-independent
analytical solution in the deep BEC regime. The larger $d$ case
converges faster.

In Fig.~\ref{fig:BEClimit}(b), we present a semi-log plot of $-\mu$
and $\Delta$ as a function of $1/k_Fa$ for $t=0.05E_F$ and $k_Fd=2$,
and compare the fully numerical solutions (solid lines) with the
analytic expressions (dashed lines) given by Eqs.~(\ref{eq:mu}) and
(\ref{eq:Delta}). As can be readily seen for the present case, the
analytical expressions become a very good approximation for the fully
numerical solutions for $1/k_Fa > 3$.

From Fig.~\ref{fig:BEClimit}, we demonstrate that $T_c$ scales
proportionally with $k_Fa = (1/k_Fa)^{-1}$ in the BEC regime,
following Eq.~(\ref{eq:TcBEC}). This is different from its
counterpart relation, $T_c \sim 1/U$, in a pure 3D lattice
\cite{NSR,Chen1}. While the general trend is the same, however, one
does not have $1/k_Fa \propto U$ in the strong coupling limit.

The parameters $t$ and $d$ are the decisive factors for the shape of
the Fermi surface. This can be seen from that of the lattice component
in Fig.~1 of Ref.~\cite{Zhang17SR}. When $t$ is small, the first BZ of
the lattice dimension will be fully occupied. In this case, a small
$d$ means a large phase space $\pm \pi/d$ in the lattice direction,
and therefore, will bring down the Fermi level as more particles now
occupy the small $k_\parallel$ but large $|k_z|$ states. On the
contrary, a large $d$ will compress the phase space region between
$\pm\pi/d$, and thus will push up the (in-plane) Fermi level. This can
be understood from the real space perspective as well. As $d$
increases, the spacing between neighboring planes
increases. Therefore, the area density within each plane has to
increase accordingly in order to keep the overall average 3D density
fixed. In this way, the Fermi level will be pushed up to
$\mu_0 = \sqrt{2\pi n_{2D}}$, where $n_{2D}$ is the 2D fermion number
density per plane.  On the other hand, for a larger $t$, it may be
possible that the first BZ in the $z$ direction is not fully
occupied. The Fermi surface in the $z$ direction will allow a larger
dispersion when $t$ increases. Depending on the size of $(t,d)$, the
Fermi surface may possess a shape of an ellipsoid, a disc, a cylinder,
or something in between. Except for the ellipsoid, all other types of
Fermi surfaces are open. A van Hove singularity will appear at the
Fermi level at the topological transition point between open and
closed Fermi surfaces.

\begin{figure}
\centerline{\includegraphics[clip,width=3.3in]{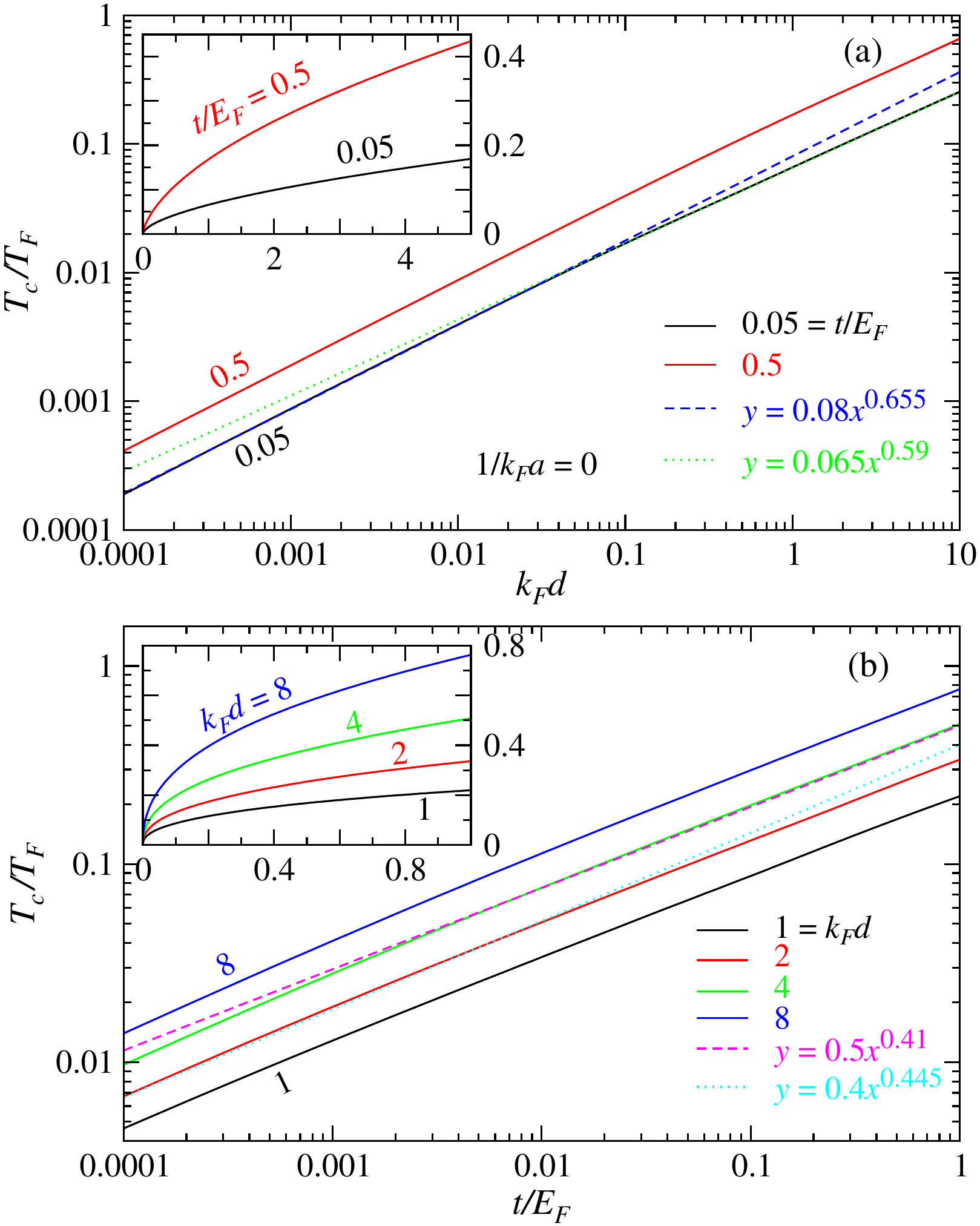}}
\caption{Behavior of $T_c$ at unitarity as a function of (a) $k_Fd$
  with $t/E_F=0.05$ (black lines) and 0.5 (red lines) and of (b)
  $t/E_F$ with $k_Fd=1$ (black), 2 (red), 4 (green) and 8 (blue
  lines), respectively. Also shown in (a) are simple power laws, which
  fit the small (blue dashed) and large (green dotted) $d$ ranges
  well, respectively. Similarly, the cyan dashed and magenta dotted
  lines are simple power laws which fit the $T_c$ curves well in the
  large and small $t$ regimes, respectively.}
\label{fig:p0inva0}
\end{figure}

Now we study the effect of $t,d$ on the behavior of $T_c$. First, we
focus on the $T_c$ behavior at unitarity as a function of $t$ and $d$,
since the unitary limit is a special point where the scattering length
diverges, and thus the system may exhibit some universal behaviors.

Shown in Fig.~\ref{fig:p0inva0} are log-log plots of $T_c$ as a
function of (a) $d$ and (b) $t$, respectively. Their linear plots are
given in the corresponding insets. Here we treat $t$ and $d$ as
independent parameters, so that they may enter the experimentally
inaccessible regime. Panel (a) covers a broad range of the $(t,d)$
parameter space, from large $t=0.5E_F$ to small $t=0.05E_F$, and from
tiny $k_Fd=0.0001$ to large $k_Fd = 10$. Surprisingly, $T_c$ exhibits
a very good power law across such a big parameter space, with a
scaling $T_c \propto d^{\alpha}$, where $\alpha$ is close to 0.655 for
small $d$ and 0.59 for large $d$. Similarly, panel (b) also covers from
$k_Fd=1$ to 8, and from $t/E_F=0.0001$ to 1.0, and $T_c$ scales as
$T_c\sim t^\beta$, where $\beta =0.445$ for small $t$ and 0.41 for
large $t$. Overall, at unitarity, we have
\begin{equation}
  T_c\sim d^\alpha t^\beta, \quad\quad \alpha = 0.59\sim 0.655,\,\,
  \beta = 0.41\sim 0.445\,.
\end{equation}

\begin{figure}
\centerline{\includegraphics[clip,width=3.3in]{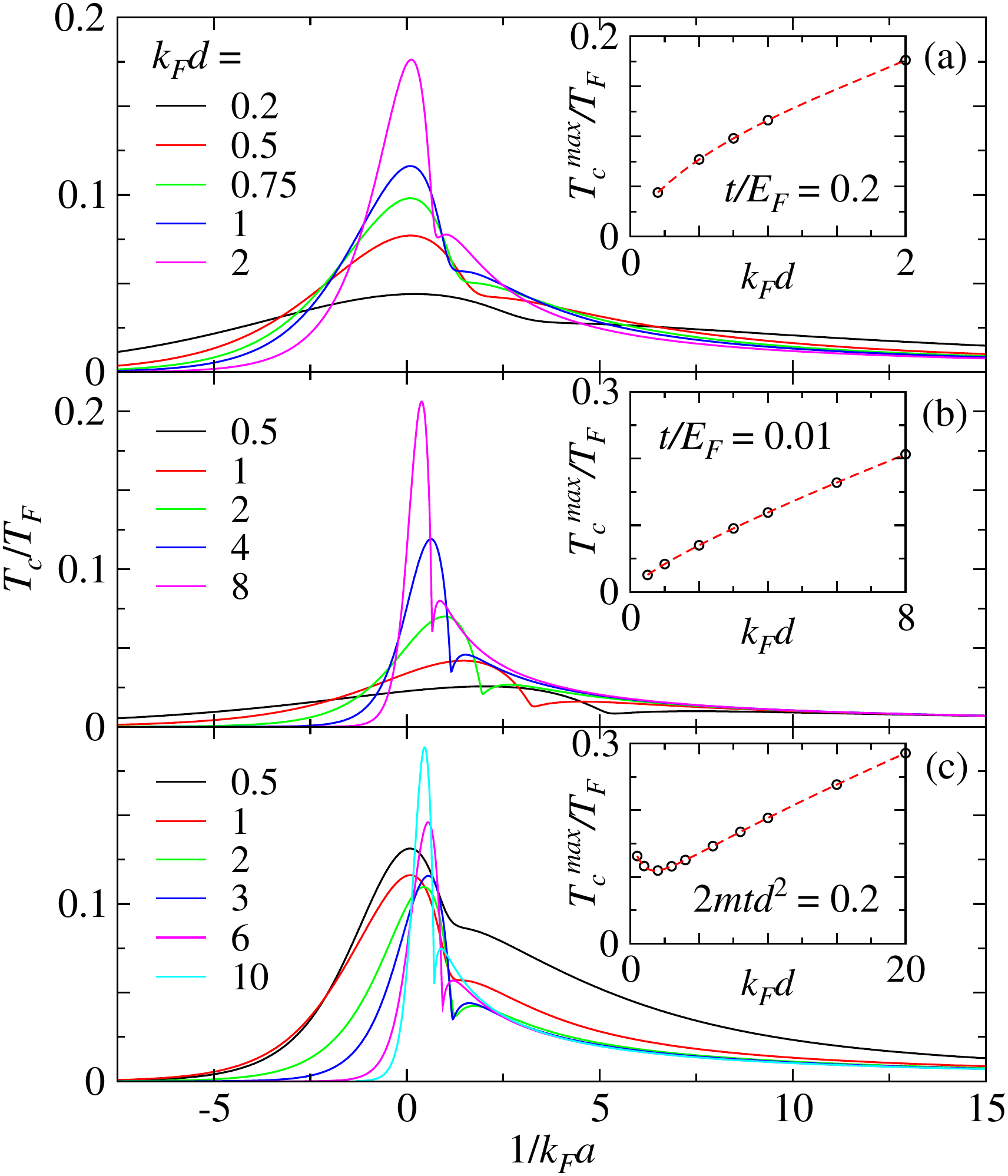}}
\caption{$T_c$ curves as a function of $1/k_Fa$ for different values
  of $k_Fd$ at fixed (a) $t/E_F=0.2$, (b) 0.01, and (c)
  $2mtd^2 = 0.2$. The maximum $T_c$ near unitarity, $T_c^\text{max}$, as a
  function of $d$, is plotted in the respective inset.}
\label{fig:p0t0.2}
\end{figure}

Next, we show in Fig.~\ref{fig:p0t0.2} the  $T_c$ curves throughout
the entire BCS-BEC crossover as a function of $1/k_Fa$ for different
$t$ and $d$. Shown in panels (a) and (b) is $T_c$ for fixed
$t/E_F=0.2$ and 0.01, respectively, but with different values of
$d$. Here we keep the product $td^2 < 1/2m$. As we can see, for fixed
$t$, the maximum $T_c$, $T_c^\text{max}$, increases with increasing $d$. At
the same time, the entire $T_c$ curve is compressed horizontally
towards unitarity, as $d$ increases. This is in accord with the
exponential behavior of $\mu \sim -e^{d/a}$ in the BEC regime. We plot
$T_c^\text{max}$ versus $d$ in the corresponding insets, which exhibits a
quasi-linear behavior. The comparison between panels (a) and (b) for
the same $d$ reveals that $T_c$ increases with $t$. Indeed, $T_c$ will
be suppressed logarithmically to zero as $t$ approaches 0
\cite{Chen1}. We also note that the peak of maximum $T_c$ moves away
from unitarity towards the BEC side as $d$ decreases. This also has to
do with the exponential behavior of $\mu \sim -e^{d/a}$.

In Fig.~\ref{fig:p0t0.2}(c), we present the $T_c$ curves for fixed
$2mtd^2 = 0.2$ while changing $k_Fd$ from 0.5 to 10. Since $t$
decreases as $d$ increases, it is not surprising to see nonmonotonic
behavior of $T_c^\text{max}$ versus $d$, as shown in the
inset. Nonetheless, we still see an overall increase of $T_c^\text{max}$
with $d$ while keeping $td^2$ fixed. This increase is not as dramatic
as the fixed $t$ cases, reflecting the competing effects between
increasing $d$ and decreasing $t$.

It should be pointed out that the increase of $T_c$ in
Fig.~\ref{fig:p0t0.2} will disappear if we use the respective Fermi
level $\mu_0$ in the noninteracting limit as the energy unit, as
$\mu_0$ increases with $t$ and $d$ as well \cite{UltraHighTc}. Nevertheless,
this increase does make sense when one compares $T_c$ with the 3D
homogeneous system of the same fermion density.

\begin{figure}
\centerline{\includegraphics[clip,width=3.3in]{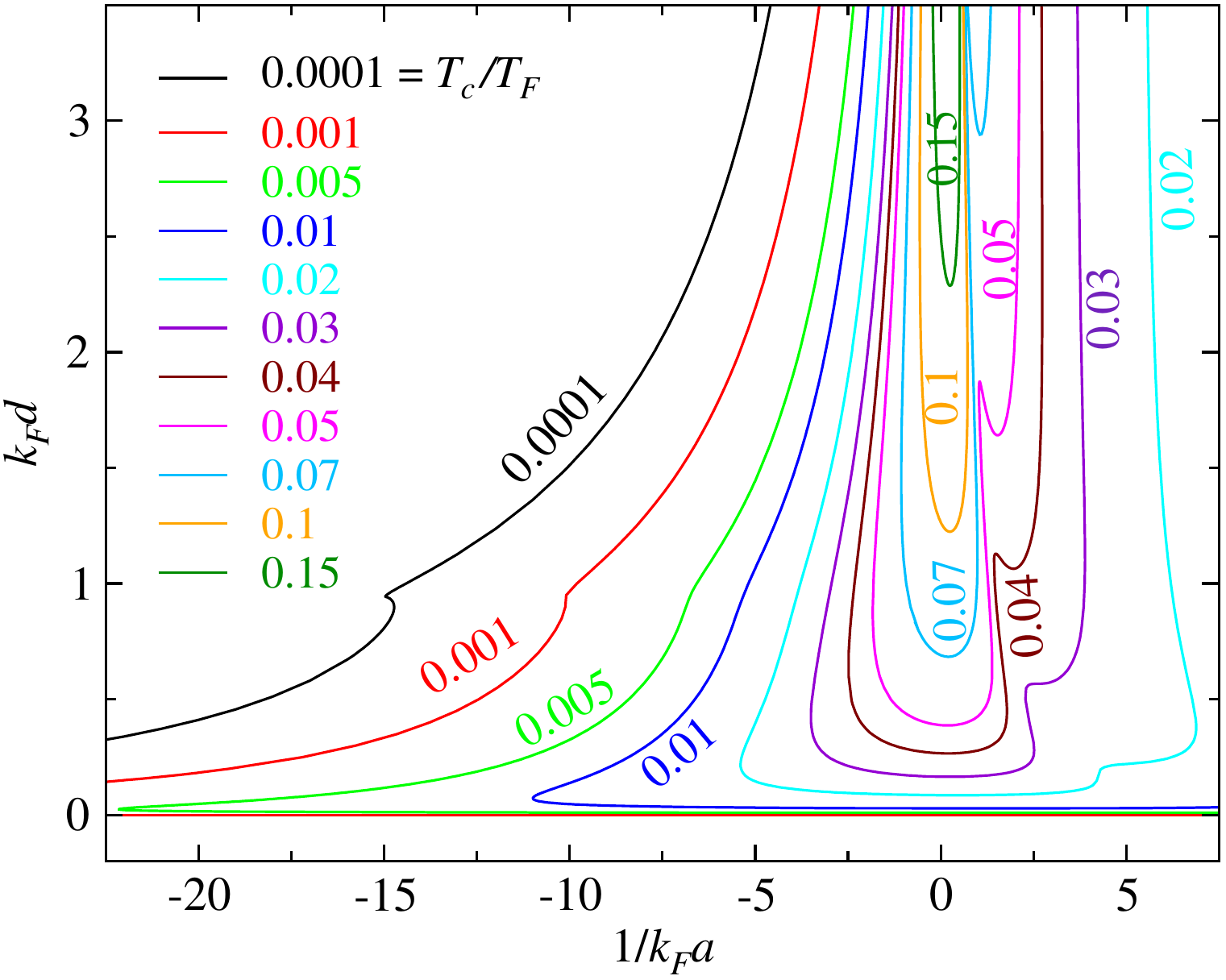}}
\caption{Contour plot of $T_c/T_F$ in the $k_Fd$ -- $1/k_Fa$ plane for fixed $t/E_F = 0.1$. The corresponding $T_c$ values are labeled near the curves. }
\label{fig:Contour}
\end{figure}

With multiple tunable parameters, the complete superfluid phase
diagram is very complex, occupying a hyper volume in the high
dimensional phase space. We can show only hypersurfaces corresponding
certain fixed parameters. As an example, presented in
Fig.~\ref{fig:Contour} are $T_c$ contours in the $k_Fd$ -- $1/k_Fa$
plane with fixed $t/E_F = 0.1$. From this figure, one can see that the
highest $T_c \gtrsim 0.15 $ is achieved at large $d$ near
unitarity. The higher concentration of curves at large $d$ indicates
that the $T_c$ curve is highly compressed towards unitarity as $d$
increases, as shown in Fig.~\ref{fig:p0t0.2}. On the contrary, when
$d$ becomes small ($\ll 1$), the $T_c$ curve as a function of $1/k_Fa$
will be suppressed down and expanded along the $1/k_Fa$ axis. One can
also consider a vertical cut at fixed $1/k_Fa$ in
Fig.~\ref{fig:Contour}. A cut at $1/k_Fa =0$ will yield a curve as in
the inset of Fig.~\ref{fig:p0t0.2}(a,b). The peak/dip structure of the
$T_c$ contours at positive $1/k_Fa$ for $k_Fd = 0.03\sim 0.07$ in
Fig.~\ref{fig:Contour} is associated with the dip near $\mu=0$ in the
$T_c$ vs $1/k_Fa$ curves, as shown in Fig.~\ref{fig:p0t0.2}. Another
feature that is worth mentioning is the small kink in the contours on
the BCS side, especially for the lowest $T_c/T_F =0.0001$. As can be
seen, for all contours, this kink happens slightly below $k_Fd =
1$. For $t/E_F = 0.1$, the topology of the Fermi surface changes from
open to closed at $k_Fd\approx 0.945$. The van Hove singularity
associated with this topological change leads to logarithmic
divergence of the density of states, and thus significantly enhances
$T_c$, so that the $T_c$ contour will deform towards weaker pairing
strength, as indeed shown by the low $T_c$ contours in
Fig.~\ref{fig:Contour}. This singularity effect is washed out
gradually by thermal broadening as $T$ increases. It becomes barely
noticeable for $T_c/T_F \geq 0.01$. Note that the van Hove singularity
effect on $T_c$ cannot readily be seen in other types of plot.

\subsection{Gaps in the superfluid phase }

\begin{figure*}
\centerline{\includegraphics[clip,width=5.4in]{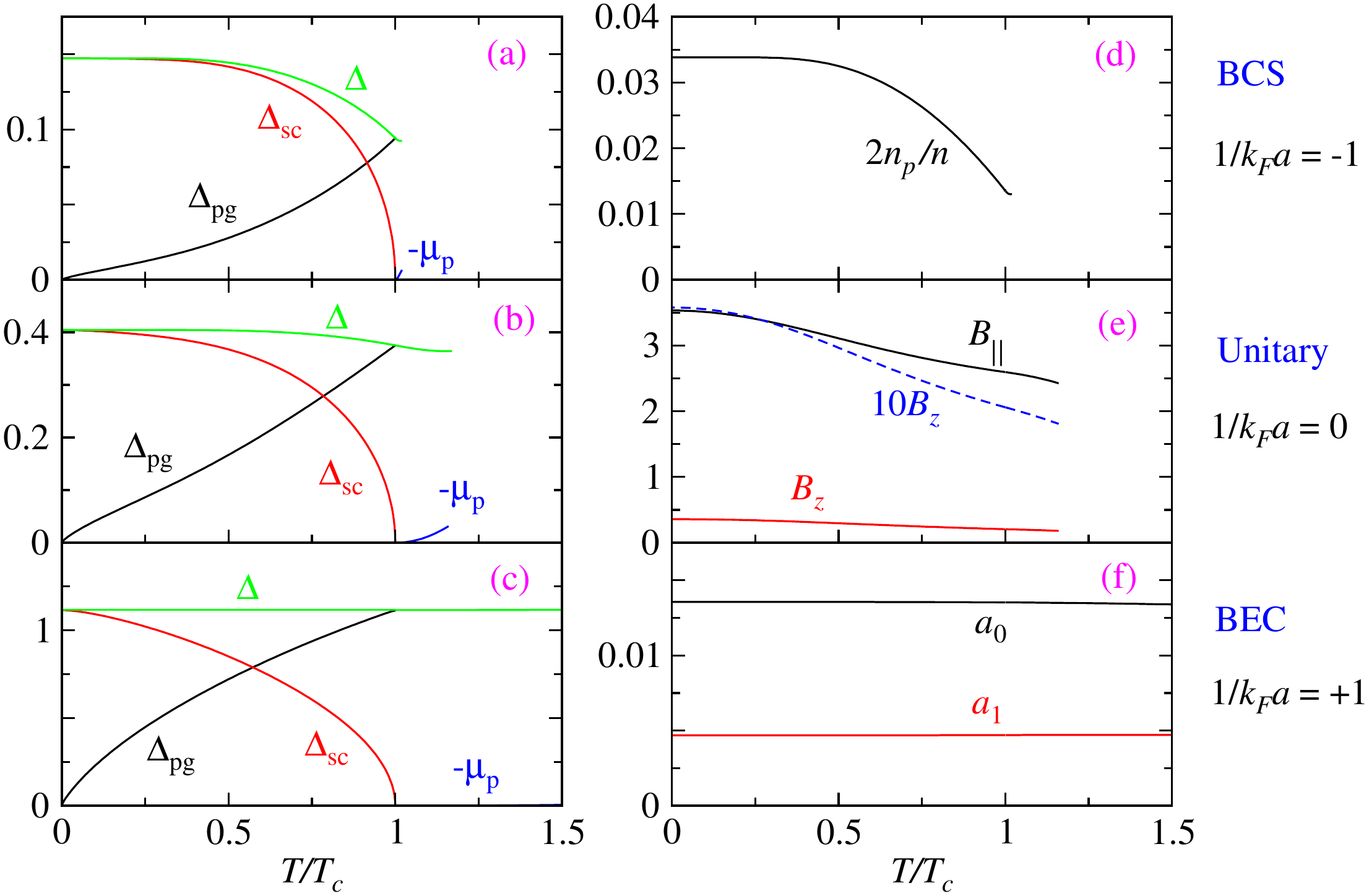}}
\caption{Behavior of the gaps and $-\mu_p$, as labeled, as a function
  of $T/T_c$, for (a) $1/k_Fa = -1$, (b) 0, (c) +1, with
  $T_c/T_F=0.07060$, 0.13135 and 0.05156, for the BCS, unitary, and
  BEC regimes, respectively. Also plotted are (d) $2n_p/n$ for
  $1/k_Fa = -1$, (e) $B_\parallel$ and $B_z$ as well as $10 B_z$ for
  $1/k_Fa = 0$, and (f) $a_0$ and $a_1$ for $1/k_Fa = +1$. Here
  $t/E_F = 0.1$, and $k_Fd=2$.  Gaps and chemical potential are in
  units of $E_F$. The coefficients $B$'s are in units of $1/2m$, $a_0$
  and $a_1$ in units of $k_F^3/E_F^2$ and $k_F^3/E_F^3$,
  respectively. }
\label{fig:Gaps}
\end{figure*}

In Fig.~\ref{fig:Gaps}, we present, as an example, the behavior of the
order parameter $\Delta_{sc}$ (red), the pseudogap $\Delta_{pg}$
(black) and the total gap $\Delta$ (green curves) and a few relevant
quantities as a function of temperature in the superfluid phase. Also
plotted is the solution slightly above $T_c$, especially for the pair
chemical potential $\mu_p$. Shown in the figure is for the case of
$k_Fd =2$, $t/E_F = 0.1$ for $1/k_Fa = -1$, 0, and +1, for the BCS,
unitary, and BEC regimes, respectively.  There exists a pseudogap in
all cases throughout the BCS--BEC crossover, as in the regular 3D
continuum case \cite{Kosztin1}. The order parameter $\Delta_{sc}$
sets in at $T_c$ with decreasing $T$, while the pseudogap
$\Delta_{pg}$ starts to decrease. The total gap increases with
decreasing $T$ in the BCS regime, where $\Delta_{pg}$ is small, but
stays roughly constant for the unitary and BEC cases. Above $T_c$, the
pair chemical potential $\mu_p$ starts to decrease from 0 with
increasing $T$. As seen in the figure, $-\mu_p$ increases much faster
in the BCS than in the BEC regimes as a function of $T$ above $T_c$.
This makes our simplified BCS form of the pseudogap self energy
become quickly less accurate above $T_c$ in the BCS regime. The
curves stop roughly where the approximation becomes inaccurate.

Figure \ref{fig:Gaps}(a) suggests that $-\mu_p$ increases linearly
with $(T-T_c)$. Indeed, as one often finds in the weak fluctuation
treatment in the framework of the mean-field BCS theory,
$\mu_p\propto -(T-T_c)$ above $T_c$ in the BCS limit
\cite{VarlamovBook,Boyack_2018}. As the pairing becomes stronger,
$\mu_p$ becomes quadratic in $(T-T_c)$, as manifested in
Fig.~\ref{fig:Gaps}(b). For the BEC case in Fig.~\ref{fig:Gaps}(c),
$-\mu_p$ stays small up to very high $T\gg T_c$. In this case, the
gaps are large, and essentially all atoms form pairs, so that the
system exhibits behaviors that are close to an ideal Bose gas.

We show in Fig.~\ref{fig:Gaps}(d) the pair fraction for the BCS case,
where the pairing is weak and the pair fraction is small. The
temperature dependence of $n_p$ follows roughly that of $\Delta^2$ via
Eq.~(\ref{eq:PG}), as $a_0$ is less sensitive to $T$. The pair density
$n_p$ increases with $1/k_Fa$ and becomes $n/2$ for $1/k_Fa = +1$,
which has $\mu/E_F \approx -0.12 < 0$ for all $T\leq T_c$. For
$B_\parallel$ and $B_z$, we show for the unitary case in
Fig.~\ref{fig:Gaps}(e). Their temperature dependencies are stronger in
the BCS regime and weaker in the BEC regime. In addition,
$B_\parallel$ approaches $1/4m$ in the BEC limit. At the same
time, $B_z$ becomes exponentially smaller in the BEC regime, as given
by Eq.~(\ref{eq:BzBEC}). Finally, the $T$ dependencies of $a_0$ and
$a_1$ are shown in Fig.~\ref{fig:Gaps}(f) for the BEC case. Both $a_0$
and $a_1$ become essentially $T$ independent, as does the total
gap. It is also evident that $a_1 T_c \ll a_0$ for this case. The
$a_1$ term in the inverse $T$-matrix expansion is quantitatively
important only in the BCS regime, where we find
$a_1 T_c / a_0 \sim 10$ for the case in Fig.~\ref{fig:Gaps}(a).
More detailed discussions of the influence of the $a_1$
term can be found in Ref.~\cite{ChenPhD} for the somewhat similar 3D
continuum case.

\subsection{Superfluid density }

\begin{figure*}
\centerline{\includegraphics[clip,width=5.in]{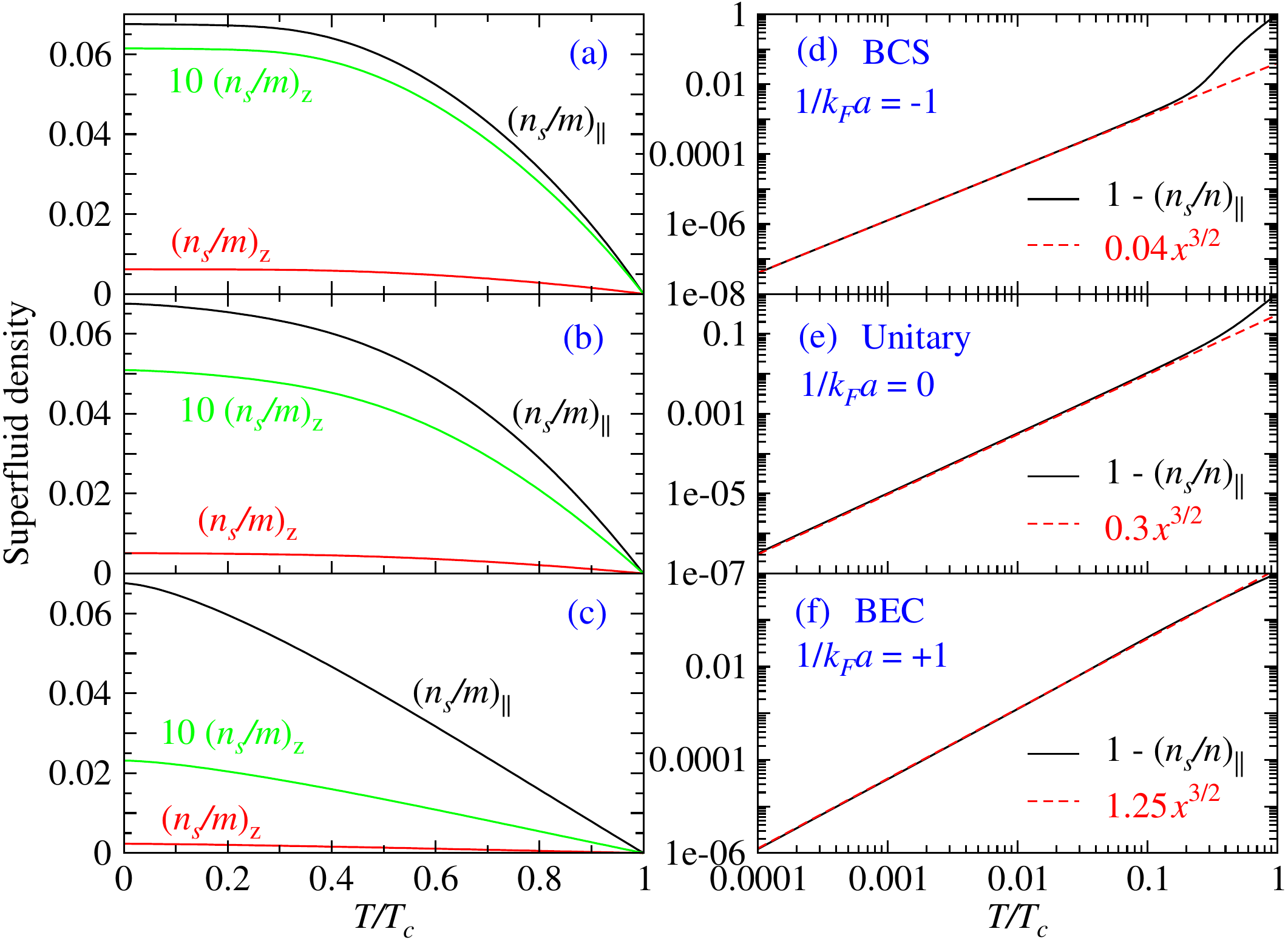}}
\caption{Behavior of the in-plane (black curves) and lattice
  components (red curves) of the superfluid densities, as labeled, as
  a function of $T/T_c$, for (a) $1/k_Fa = -1$, (b) 0, (c) +1, for the
  BCS, unitary, and BEC regimes, respectively, corresponding to
  Fig.~\ref{fig:Gaps}. Also shown is $10 (n_s/m)_z$ (green curves) for
  clarity. Plotted in (d-f) are the corresponding normal fluid fraction
  $1-(n_s/n)_\parallel$ (black solid curves) in log-log scales, and
  simple $T^{3/2}$ power laws (red dashed lines) for comparison.  The
  zero $T$ value of $(n_s/m)_\parallel$ is given by $2/3\pi^2$ in our
  convention of units.  }
\label{fig:Ns}
\end{figure*}

Now we present the result for the superfluid density
calculations. Shown in Fig~\ref{fig:Ns} are the in-plane and lattice
components of the superfluid density, from top to bottom, for the BCS,
unitary and BEC cases (of Fig.~\ref{fig:Gaps}), respectively. The left
column presents $(n_s/m)_\parallel$ (black) and $(n_s/m)_z$ (red), as
well as $10 (n_s/m)_z$ (green curves). The right column presents the
in-plane normal fluid fraction, $1-(n_s/n)_\parallel$, as a function
of $T/T_c$ in log-log scales (black solid lines). For comparison, we
plot simple power laws of $(T/T_c)^{3/2}$ (red dashed lines), with
different coefficients to fit roughly the corresponding solid
lines. Here the message is clear. In the BCS case, the linear plot in
Fig~\ref{fig:Ns}(a) looks very much like an exponential $T$ dependence
at low $T$. Only a log-log plot in Fig~\ref{fig:Ns}(d) reveals that
the leading dominant term is actually a $T^{3/2}$ power law. The small
coefficient, 0.04, in front of $(T/T_c)^{3/2}$, is consistent with the
flatness of $(n_s/m)_\parallel$ at low $T$ in
Fig~\ref{fig:Ns}(a). Nevertheless, the power law contributions from
finite momentum pairs always dominate the exponentially activated term
from Bogoliubov quasiparticles. As the pairing strength, or $1/k_Fa$,
increases, the magnitude of the power law term becomes larger. For the
unitary case, even in the linear plot in Fig~\ref{fig:Ns}(b),
$(n_s/m)_\parallel$ deviates strongly from exponential behavior. The
coefficient increases to 0.3, as shown in Fig.~\ref{fig:Ns}(e). For
the BEC case in Fig~\ref{fig:Ns}(c), the quasiparticle contributions
become negligible, and $(n_s/m)_\parallel$ becomes almost purely a
$(T/T_c)^{3/2}$ power law. As one can see in Fig~\ref{fig:Ns}(f), the
dash line overlays with the solid curve essentially for the entire range
of $T\le T_c$. It should be noted, however, that the coefficient is
now 1.25, larger than 1. This reflects the fact that the system is
quasi-2D rather than 3D; a pure $(T/T_c)^{3/2}$ is only for a pure 3D
case. Indeed, as one can see from Fig~\ref{fig:Ns}(c),
$(n_s/m)_\parallel$ becomes more of a straight line for the upper half
of $T/T_c$, to be compatible with the larger-than-unity coefficient
1.25. Theoretically, as $T$ becomes higher, more high in-plane
momentum $q_\parallel$ pairs will be excited, which can feel the
quasi-two dimensionality.

The lattice component of the superfluid density, $(n_s/m)_z$, (red
curves) in Fig~\ref{fig:Ns}(a-c) is substantially smaller than
$(n_s/m)_\parallel$, as discussed earlier. Its temperature dependence
is close to that of $(n_s/m)_\parallel$, as can be seen more clearly
from the (green) $10\times$ magnified curves. This is because both are
mainly governed by the prefactor
$\Delta_{sc}^2 = \Delta^2 -\Delta^2_{pg}$ in Eq.~(\ref{eq:Ns}), and
the second term, $\Delta_{pg}^2 \propto T^{3/2}$, yields the $T^{3/2}$
power law for both components of $(n_s/m)$. For the present $s$-wave
pairing, the rest of Eq.~(\ref{eq:Ns}) yields an exponential $T$
dependence for the normal fluid, $e^{-\Delta/T}$, and thus becomes
negligible at low $T$.  It is also evident that $(n_s/m)_z$ decreases
as the pairing becomes stronger toward BEC. This can be understood
since $\vk^2$ becomes more widespread in momentum space as $1/k_Fa$
increases, and thus pairs feel more strongly the effect of lattice
momentum cutoff in the $z$ direction, so that the system becomes
effectively more 2D. On the other hand, the mobility of the pairs is
controlled by $t_B$, which decreases rapidly with $1/k_Fa$. This
determines the magnitude of $(n_s/m)_z$ in the BEC regime.

\section{Conclusions}

In summary, we have studied the ultracold atomic Fermi gases in a 1D
optical lattice with a pairing fluctuation theory, as they undergo a
BCS-BEC crossover. We find that $T_c$ decreases with $1/k_Fa$ in the
BEC regime and approaches asymptotically $T_c/T_F = \pi an/2m$, which
is independent of the lattice parameters $t$ and $d$. Both $|\mu|$ and
the gap $\Delta$ grow exponentially as $e^{d/a}$ and $e^{d/2a}$,
respectively, in the BEC regime so that the pair hopping integral
$t_B$ decreases as $e^{-d/a}$. Moreover, the (maximum) $T_c$ near
unitarity increases with both $t$ and $d$, with fractional power law
exponents. On the BCS side, the effect of van Hove singularity on
$T_c$ has been identified in the $T_c$ contours.

We find generally a pseudogap above and below $T_c$, away from the
extreme BCS limit. While the total gap $\Delta$ is a smooth function
across $T_c$, the order parameter sets in at $T_c$, and the pseudogap
starts to decrease as $T$ decreases below $T_c$. Our calculated
behavior of the pair chemical potential $\mu_p$ above $T_c$ are also
in good agreement with existing literature.  At low $T$,
$\Delta_{pg}^2 \sim T^{3/2}$. This leads to $T^{3/2}$ power laws for
the low $T$ dependence of the superfluid density, despite that it
looks visually like exponential in the BCS regime.

Our findings have not been reported in the literature. Although
precise control and measurements of the gaps and superfluid density
remains challenging experimentally at present, we believe that our
predictions can be tested in future experiments.

\section{Acknowledgments}
We thank the useful discussions with Chenchao Xu, Lin Sun and Yanming
Che. This work was supported by the NSF of China (Grant No. 11774309
and No. 11674283), and the NSF of Zhejiang Province of China (Grant
No. LZ13A040001).  C. Lee was supported by the NSF of China (Grants
No. 11874434 and No. 11574405).

\appendix

\section{Coefficients of the Taylor-expanded inverse $T$-matrix}

In this Appendix, we present concrete expressions for the
coefficients of the Taylor expansion of the inverse $T$-matrix,
$t(\Omega,\mathbf{q})$, after analytical continuation,
\begin{equation} 
t^{-1}_{pg}({\mathbf{q}}, \Omega) = a_1\Omega^2 + a_0 (\Omega -\Omega_\mathbf{q}
+ \mu_p +i \Gamma_{{\mathbf{q}}, \Omega})\;.  
\label{Omega_q:exp}
\end{equation}
Here $\mu_p = t^{-1}(0,0)/a_0$, which vanishes for $T\leq T_c$.
In the long wavelength limit, 
\begin{equation}
\Omega_{\mathbf{q}} = B_\parallel q_\parallel^2 + B_z q_z^2 
 \equiv \frac{q_\parallel^2} {2 M_{\parallel}} 
+ \frac{q_z^2} {2 M_z}\,, 
\end{equation}
with $B_z =   t_B d^2$.

Before expansion, the inverse $T$ matrix is given by
\begin{eqnarray}
t^{-1}_{\mb{q},\Omega+i0^+} &=& U^{-1} + \sum_\mb{k} \left[
                                \frac{1-f(\Ek)-f(\ekq)} {\Ek + \ekq - \Omega-i0^+} u_\mb{k}^2 \right. \nonumber\\
  &&{} \left. -
  \frac{f(\Ek) - f(\ekq)} {\Ek - \ekq + \Omega+i0^+} v_\mb{k}^2 \right]
\;.  
\label{Omega_q:t-exp}
\end{eqnarray}
Then we have
\begin{eqnarray} 
a_0 & = & \frac{1}{2\Delta^2}\sum_\mb{k} \bigg[ [1-2f(\ek)] -
  \frac{\ek}{\Ek} [1-2f(\Ek)]\bigg] \nonumber\\
   &  = & \frac{1}{2\Delta^2} \bigg[ n- 2\sumk f(\ek)\bigg] \;,\\
\label{Omega_q:a0}
a_1 &=&\frac{1}{2\Delta^4} \sumk \Ek \left[ \left( 1+\frac{\ek^2}{\Ek^2}
        \right) [1-2f(\Ek)]\right.\nonumber\\
  &&{}\left.- 2 \frac{\ek}{\Ek}[1-2f(\ek)] \right] 
\label{Omega_q:a1}
\end{eqnarray}
and the imaginary part
\begin{eqnarray}
\Gamma_{\mb{q},\Omega} &=&\! \frac{\pi}{a_0} \sumk \Big\{
 \!\left[1\!-\!f(\Ek)\!-\!f(\ekq)\right] \uk^2 \delta (\Ek\!+\!\ekq-\!\Omega)
  \nonumber\\ 
&&{}\!\! +  \left[f(\Ek)\!-\! f(\ekq)\right]\! \vk^2
  \delta(\Ek\!-\!\ekq\!+\!\Omega)\Big\}\:.
\label{Omega_q:Gamma_q}
\end{eqnarray}
We have $\Gamma_{\mb{q},\Omega}=0$ when
$-(\Ek-\ekq)_\text{min} < \Omegaq < (\Ek+\ekq)_\text{min}$, and in
general $\Gamma_{\mb{q},\Omega}$ is much smaller than $\Omegaq$ for
small ${q}$ at $T\leq T_c$. For details, see Ref.~\cite{ChenPhD}.

The pair dispersion coefficients are given by
\begin{widetext}
\begin{eqnarray}
 B_i &=& \left.\frac{1}{2}\frac{\partial^2\Omegaq}{\partial q_i^2}\right|_{\mb{q}=\mb{0}}\nonumber\\
 & =&  -\frac{1}{2a_0\Delta^2} \sumk \left\{ \left[2f^\prime(\ek) 
  + \frac{\Ek}{\Delta^2} 
   \left[ \left( 1+\frac{\ek^2}{\Ek^2}
      \right) [1-2f(\Ek)] \right. - 2 \frac{\ek}{\Ek}[1-2f(\ek)] \right] \right]
 \left( \frac{\partial\ek}{\partial k_i}\right)^2 \nonumber \\
 &&{}\left. - \frac{1}{2} \left[ [1-2f(\ek)] - \frac{\ek}{\Ek} 
   [1-2f(\Ek)] \right] \frac{\partial^2 \ek}{\partial k_i^2} 
    \right\} \;.
\label{Omega_q:expression}
\end{eqnarray}
\end{widetext}
Given the dispersion
$\ek = \dfrac{k_\parallel^2}{2m} -2t [1-\cos (k_zd)] -\mu$ for 1DOL,
we have, for $i=x,y$,
\[
\left(\frac{\partial\ek}{\partial k_i}\right)^2 = \frac{k_i^2}{m^2} \;, 
\qquad
\frac{\partial^2\ek}{\partial k_i^2} = \frac{1}{m}\;,
\]
and for $i=z$,
\[
\left(\frac{\partial\ek}{\partial k_z}\right)^2 = (2td)^2\sin^2(k_zd) \;, 
\quad
\frac{\partial^2\ek}{\partial k_z^2} = 2td^2\cos (k_zd)\;.
\]



%

\end{document}